\documentclass[a4paper,11pt]{article}
\pdfoutput=1 

\usepackage{jcappub} 

\usepackage[T1]{fontenc} 
\usepackage{titlesec}
\usepackage{subfig}

\title{\boldmath Let Effective Field Theory of Inflation flow: stochastic generation of
models with red/blue tensor tilt}

\author[a,b,c]{Giulia Capurri,}
\author[d,e,f]{Nicola Bartolo,}
\author[g,h,i]{Davide Maino}
\author[d,e,f,j]{and Sabino Matarrese}

\affiliation[a]{SISSA, Astrophysics and Cosmology Sector, via Bonomea 265, 34136,
Trieste, Italy}
\affiliation[b]{INFN, Sezione di Trieste, Via Valerio 2, 34127 Trieste, Italy}
\affiliation[c]{IFPU, Institute for Fundamental Physics of the Universe, via Beirut 2, 34151, Trieste, Italy}
\affiliation[d]{Dipartimento di Fisica e Astronomia ``Galileo Galilei'', Universit\`a  di Padova, 35131 Padova, Italy}
\affiliation[e]{INFN, Sezione di Padova, 35131 Padova, Italy}
\affiliation[f]{INAF - Osservatorio Astronomico di Padova, I-35122 Padova, Italy}
\affiliation[g]{Dipartimento di Fisica, Universit\`a degli Studi di Milano, Via Celoria, 16, Milano, Italy}
\affiliation[h]{INAF/IASF Milano, Via E. Bassini 15, Milano, Italy}
\affiliation[i]{INFN, Sezione di Milano, Via Celoria 16, Milano, Italy}
\affiliation[j]{Department of Theoretical Physics and Center for Astroparticle Physics (CAP), CH-1211
Geneva 4, Switzerland}

\emailAdd{giulia.capurri@sissa.it}
\emailAdd{nicola.bartolo@pd.infn.it}
\emailAdd{davide.maino@mi.infn.it}
\emailAdd{sabino.matarrese@pd.infn.it}

\abstract{We extend the method of Flow Equations to the Effective Field Theory framework of inflation, in order to investigate the observable predictions of a very broad class of inflationary models. Focusing our attention on the gravitational-wave sector, we derive a general expression for the consistency relation for effective models and provide a numerical implementation which allows to study how the generated models populate the $(r,n_{t})$ plane. We analyse $5 \times 10^{4}$ realizations of inflationary scenarios that respect the Null-Energy Condition ($\epsilon > 0$) and $5 \times 10^{4}$ realizations that violate it ($\epsilon < 0$). In both cases, $90 \%$ of the viable models are below the most recent upper bound on the tensor-to-scalar ratio from {\it Planck} and BICEP2/Keck Array BK15  data: $r_{0.002} < 0.056$ at 95 $\%$ CL. We find that general EFT inflationary models with $\epsilon > 0$ are typically characterized by $n_{t} < 0$, whereas the vast majority of NEC-violating models presents a blue-tilted spectrum ($n_{t} > 0$). Since a blue tensor spectral index implies more power on small scales, this result is of considerable interest in view of a possible direct detection of the primordial gravitational-wave background.}

\newcommand{\mpl}{M_{\textrm{pl} }}
\newcommand{\mpls}{M_{\textrm{pl}}^{2}}
\newcommand{\integral}{\int d^{4}x \sqrt{-g}}
\newcommand{\tens}{_{\textrm{\tiny{T}}}}
\newcommand{\q}{_{\textrm{\tiny{Q}}}}
\newcommand{\m}{\textrm{\tiny{M}}}

\begin{document}
\maketitle
\flushbottom


\section{Introduction}
\label{sec:intro}
Inflation \citep{PhysRevD.23.347} is the most accredited theory of the early universe. It yields a natural solution to the the so-called horizon, flatness and monopoles problems \citep{Linde:1981mu} and provides a causal mechanism for the generation of primordial density perturbations. Moreover, it has been shown to be in perfect agreement with all the observations made so far \citep{Akrami:2018odb}. Despite its success, inflation is far from being thoroughly characterized, both from the theoretical and observational points of view. Throughout the years, many different inflationary models have been proposed and the advent of new generation experiments, mainly focused on the detection of B-modes in the Cosmic Microwave Background \citep{Hazumi:2019lys, Ade:2018sbj, Abazajian:2016yjj}, carries on the effort to test inflation. A general prediction of inflation is in fact the generation of a stochastic background of primordial gravitational waves, the detection of which would be of extreme importance for cosmology. Not being expected in the framework of non-inflationary early universe models, primordial gravitational waves constitute a smoking-gun probe of inflation \citep{Guzzetti:2016mkm}.
In this complex scenario, it is important to characterize the most general theory of inflation and to develop tools for studying the general observable predictions of inflation, without relying on a single specific model. The former can be addressed by the Effective Field Theory (EFT) of inflation \citep{Creminelli:2006xe}, which provides the most general theory describing the fluctuations around a simple slow-roll background, in the case of single-field models.~\footnote{For a similar approach see~\cite{Weinberg:2008hq}.}  For what concerns the latter issue, on the other hand, an alternative approach is to seek to generate inflationary models via some stochastic process, exploiting numerical techniques where appropriate. The Flow Equations \citep{Hoffman:2000ue, Kinney:2002qn} are the archetypal of such methods.
In this work, we derive an extended version of the Flow Equations that can be applied to the effective inflationary action, in order to stochastically generate models in the EFT framework and to compute their observable predictions. Since primordial tensor perturbations encode unique information about the physics of inflation, we focus our attention on the gravitational-wave sector. We therefore derive a general consistency relation (linking the 
tensor-to-scalar perturbation $r$ to the tensor spectral index $n_t$) valid for EFT models and we study the features of the tensor power-spectra of a large number of simulated models. In particular, we find that the majority of the generated points have a tensor-to-scalar ratio compatible with the {\it Planck}/BICEP2/Keck Array BK15 95$\%$ CL upper limit $r_{0.002} < 0.056$ \citep{Akrami:2018odb,Ade:2018gkx}. Moreover, we find that EFT models that respect the Null-Energy Condition (NEC) are characterized by a red tensor spectrum ($n_{t} < 0$), whereas EFT models that do not respect NEC present a blue tensor tilt ($n_{t} > 0$).

The paper is organised as follows. In section \ref{sec:eft_flow}, after a rapid recap of the method of Flow Equations and of the EFT of inflation, we derive the system of Flow Equations for the EFT framework. In section \ref{sec:observable_predictions}, we derive a general form of the consistency relation for the considered models, we briefly describe the algorithm for the integration of the Flow Equations and we discuss how $5 \times 10^{4}$ generated models populate $(r,n_{t})$ plane in both scenarios, where NEC is or is not respected. Finally, in section \ref{sec:conclusion} we outline our conclusions and future prospects.


\section{Flow Equations for EFT models}
\label{sec:eft_flow}
The main theoretical result of this work is the derivation of a system of Flow Equations for the generation of stochastic realizations of inflationary models, picked up from the very broad class of second-order EFT models. First though, it is worth to recall some of the basic features about the method of the Flow Equations and of the EFT approach to inflation.

\subsection{The method of Flow Equations} 
\label{subsec:flow_eqs}
The Hubble Flow Equations \citep{Hoffman:2000ue,Kinney:2002qn,Ramirez:2005cy} (see also \citep{Liddle:2003py, Vennin:2014xta}) were first introduced for standard single-field slow-roll inflationary models. For those models, it is straightforward to define an infinite hierarchy of slow-roll parameters that completely specify the action starting from the Hubble parameter $H$ and its derivatives with respect to the field $\phi$ \citep{Liddle:1994dx}. The Hubble Flow Equations are a system of coupled differential equations whose solutions give the evolutionary paths of the slow-roll parameters and that can be used to generate many inflationary models. 
Another important property is that the main observable quantities related to inflation, namely the spectral indices $n_{s}$ and $n_{t}$ and the tensor-to-scalar ratio $r$, can be expressed in terms of the slow-roll parameters up to an arbitrary order \citep{Stewart:1993bc}. Therefore, solving numerically the truncated system of Hubble Flow Equations for suitably defined initial conditions allows to calculate the observable predictions of a large number of inflationary models, without relying on the explicit form of the action \citep{Peiris:2003ff,Peiris:2008be}. Conversely, this method can be applied to observed and simulated data in order to reconstruct the potential $V(\phi)$ \citep{Easther:2002rw,Kinney:2003uw,Peiris:2006ug}. 

The method of Flow Equations was also extended to more complex inflationary scenarios, i.e. models with non-standard kinetic term \citep{Peiris:2007gz,Powell:2008bi,Bessada:2009pe} or with an even more general action of the type $\mathcal{L}(X, \phi)$ (where $X = - \frac{1}{2} g^{\mu\nu} \partial_{\mu}\phi \partial_{\nu} \phi$, see~\cite{Chen:2006nt}) by  \citep{Bean:2008ga,Agarwal:2008ah}, where $X = - \frac{1}{2} g^{\mu\nu} \partial_{\mu}\phi \partial_{\nu} \phi$. An important remark is that  for the these more complex models more than one hierarchy of slow-roll parameters is needed, in order to completely specify the action. For example, models with non-standard kinetic term are characterized by a speed of propagation of scalar perturbations $c_{s} \neq 1$. It can be shown that a further slow-roll hierarchy built from $c_{s}$ and its derivatives with respect to $\phi$ is needed, in order to fully specify the action of these models. This will be very important for our purposes, since, as we will see in the next paragraph, the effective action that we want to reconstruct with a suitable system of Flow Equations is composed of many parameters. 
 
\subsection{The Effective Field Theory of inflation}
\label{subsec:eft}
The EFT of inflation \citep{Creminelli:2006xe,Cheung:2007st,Piazza:2013coa,Weinberg:2008hq} is a very general approach able to describe fluctuations around a quasi-de Sitter background, in the case of single-field models. In characterizing a wide variety of possible high-energy corrections to simple slow-roll inflation models, the EFT formalism allows a unified approach to inflation. 
The general form of the effective action in the comoving gauge, organized in powers of the number of perturbations and in terms of increasing number of derivatives, is given by

\begin{equation} \label{eq:action_com_gauge}
\begin{aligned}
S = & \integral  \biggl[ \dfrac{1}{2} \mpls R - c(t)g^{00} -\Lambda(t) + \\
& + \dfrac{1}{2!} M_{2}(t)^{4}(g^{00}+1)^{2} + \dfrac{1}{3!} M_{3}(t)^{4}(g^{00}+1)^{3} + \dots \\
&- \dfrac{\bar{M}_{1}(t)^{3}}{2} (g^{00}+1)\delta K^{\mu}_{\mu} -  \dfrac{\bar{M}_{2}(t)^{3}}{2} \delta K^{\mu \, 2}_{\mu}
 - \dfrac{\bar{M}_{3}(t)^{3}}{2}\delta K^{\mu}_{\nu} \delta K^{\nu}_{\mu} + \dots\biggl],
\end{aligned}
\end{equation}
where $\delta K_{\mu \nu} = K_{\mu \nu} - a^{2}Hh_{\mu \nu}$ is the deviation of the extrinsic curvature of the surfaces of constant time from the unperturbed case. The first line in Eq.\eqref{eq:action_com_gauge} consists of the Einstein-Hilbert action and of the terms which contain linear perturbations around the background. All the quadratic or higher order terms are stored in the following lines. The coefficients of the linear terms, $c(t)$ and $\Lambda(t)$, are uniquely fixed by the background evolution. The higher-order terms, on the contrary, are completely free and encode all the peculiar features of the various models. In fact, there is a one-by-one correspondence between a given inflationary model and a specific linear combination of operators in the effective action obtained by turning on and off the $M$ coefficients that regulate the weight of the operators. In their turn the various $M$ coefficients (or a combination of them) turn out to be related to physical quantities that can be in principle measured and constrained. For instance, it can be shown that the propagation speed of scalar fluctuations for the $\mathcal{L}(X, \phi)$ models turns out be given in this formalism by $c_s^2=(1-2M_2^4/\mpls \dot{H})^{-1}$~\cite{Cheung:2007st}. Similarly one could compute a tensor perturbation speed as $c \tens ^{-2} = 1 - \bar{M}_{3}^{2}/\mpls$~\cite{Cheung:2007st,Noumi:2014zqa} (for specific models with $c_T \neq 1$ see, e.g.,~\cite{Kobayashi:2010cm,Kobayashi:2011nu}). 
Finally, in this work we make use of the EFT approach up to second-order in the perturbations. Despite the fact that the EFT formalism can be very efficient in deriving predictions for primordial non-Gaussian signatures, interestingly, the effective approach turns out to be very useful already at quadratic order in the perturbations, since it automatically generates an equation of motion whose solution encompasses the classical wave-function of many inflationary theories in the appropriate limits~\cite{Bartolo:2010im}. Indeed, for the power spectra, all the possible (leading-order) deviations from the standard single-field slow-roll models are systematically encoded in the size of quadratic operators.

\subsection{Results: Flow Equations in the EFT framework}
\label{subsec:flow_eft}
In order to extend the method of Flow Equations to the EFT framework, first of all we have to derive a system of equations for the slow-roll parameters that allow us to reconstruct the background dynamics, which is described by the action:

\begin{equation}
S = \integral  \biggl[ \dfrac{1}{2} \mpls R - c(t)g^{00} -\Lambda(t) \biggl].
\label{eq:EFT_bkg_action}
\end{equation}
This action is completely specified by the coefficients $c(t)$ and $\Lambda(t)$, which are fixed by the background FRW evolution of the Hubble parameter. We can therefore proceed in building the Flow Equations for the functions $H(t)$ and $c(t)$, since $\Lambda(t)$ can be easily derived from the Friedmann equations
\begin{equation} \label{eq:eft_friedmann}
H^{2}   =  \dfrac{1}{ 3\mpls}\bigl[ c(t) +\Lambda(t) \bigl] 
\qquad
\textrm{and}
\qquad
\dfrac{\ddot{a}}{a} =  - \dfrac{1}{ 3\mpls} \bigl[ 2 c(t) -\Lambda(t) \bigl].
\end{equation}
As it was the case for the standard Hubble Flow Equations and all their extended versions, we apply the Hamiltonian formalism to inflation \citep{Salopek:1990jq,Salopek:1990re,Kinney:1997ne} (see also \citep{Binetruy:2014zya,Binetruy:2016hna}) and we take as our fundamental quantity the Hubble parameter $H$ as a function of $\phi$. Comparing the time derivative of the first Friedmann equation with the continuity equation, we obtain the following expression of $\dot{\phi}$ as a function of $H(\phi)$, $c(\phi)$ and their derivatives with respect to $\phi$:
\begin{equation} \label{eq:phidot}
\dot{\phi} = - \dfrac{c(\phi)}{\mpls \, H'(\phi)}.
\end{equation} 
By means of this extremely useful equation, we can transform the time derivatives to derivatives with respect to $\phi$. In particular, the slow-roll parameter $\epsilon$ can be written as
\begin{equation}
\epsilon =  - \dfrac{\dot{H}}{H^{2}} = \dfrac{c(\phi)}{\mpls H^{2}(\phi)}.
\end{equation} 
Starting from $\epsilon(\phi)$, we can obtain the higher-order parameters via iterated derivations. Since the unknown function $c(\phi)$ appears in the expression of $\epsilon(\phi)$, we need to introduce another slow-roll parameter in order to specify the background dynamics:

\begin{equation}
\theta \equiv - \dfrac{\dot{c}}{Hc} = \dfrac{1}{\mpls} \, \dfrac{c'(\phi)}{H(\phi) H'(\phi)}.
\end{equation}
Repeatedly deriving the progenitor parameters $\epsilon(\phi)$ and $\theta(\phi)$ with respect to the number of e-folds $N$ from the end of inflation, we finally obtain the $H(\phi)$\textit{-tower}:

\begin{equation} \label{eq:h_param_eft}
\begin{aligned} 
\epsilon(\phi) & =  \dfrac{c(\phi)}{\mpls} \dfrac{1}{H^{2}(\phi)}   \\
\eta(\phi) & =  \dfrac{c(\phi)}{\mpls} \, \dfrac{H''(\phi)}{H(\phi)H'^{\,2}(\phi)} \\
&\>\>\vdots \\
^{l}\lambda(\phi)& = \biggl(\dfrac{c(\phi)}{\mpls}\biggl)^{l} \, \biggl( \dfrac{1}{H(\phi)} \biggl)^{l} \, 
\biggl( \dfrac{1}{H'(\phi)} \biggl)^{l+1} \, \dfrac{d^{l+1}H(\phi)}{d\phi^{l+1}} ,
\end{aligned}
\end{equation}
and the $c(\phi)$\textit{-tower}:

\begin{equation} \label{eq:c_param_eft}
\begin{aligned} 
\theta(\phi) & =  \dfrac{1}{\mpls} \, \dfrac{c'(\phi)}{H(\phi) H'(\phi)} \\
\kappa(\phi) & =  \dfrac{1}{\mpls} \, \dfrac{c''(\phi)}{H'^{\,2}(\phi)} \\
&\>\>\vdots \\
^{l}\xi(\phi) &= \biggl(\dfrac{c(\phi)}{\mpls}\biggl)^{l} \, \biggl( \dfrac{1}{H(\phi)} \biggl)^{l-1} \, 
\biggl( \dfrac{1}{H'(\phi)} \biggl)^{l+1} \, \dfrac{1}{c(\phi)} \, \dfrac{d^{l+1}c(\phi)}{d\phi^{l+1}} ,
\end{aligned}
\end{equation}
where $l\geq2$ and $\eta$ and $\kappa$ are $\,^{1}\lambda$ and $\,^{1}\xi$ respectively.
The Flow Equations for the previous parameters come to light through explicit calculations:

\begin{equation} \label{eq:EFT_flow_eqs_bkg}
\begin{cases}
\dfrac{d\epsilon}{dN} & =\quad  \epsilon(\theta - 2\epsilon) \\[10pt]
\dfrac{d\eta}{dN}  &=\quad  \eta(\theta - \epsilon -2\eta) +\,^{2}\lambda \\
&\, \vdots   \\
\dfrac{d\,^{l}\lambda}{dN}  &= \quad \,^{l}\lambda \bigl[l(\theta-\epsilon) - (l+1)\eta \bigl] + \,^{l+1}\lambda \\[30pt]

\dfrac{d\theta}{dN} & =\quad  \epsilon\kappa - \theta(\epsilon + \eta) \\[10pt]
\dfrac{d\kappa}{dN}  &=\quad  -2\kappa \eta +\,^{2}\xi \\
&\, \vdots   \\
\dfrac{d\,^{l}\xi}{dN}  &=\quad ^{l}\xi \bigl[(l-1)(\theta - \epsilon) - (l+1)\eta\bigl] + \,^{l+1}\xi.
\end{cases}
\end{equation}
The derivative of a slow-roll parameter of order $n$ is always of order $n+1$ in the slow-roll expansion. In particular, it is expressed as an $(n+1)$-order combination of the parameters of lower order and of a new parameter of order $n+1$. The integration of the this system of coupled equations completely specifies the dynamics of the background.

Once we have reconstructed the background dynamics, we can derive a further system of equations that will enable us to generate the evolutionary paths of the $M$ coefficients of the EFT action. As we already mentioned, particular combinations of the $M$ coefficients give birth to specific inflationary models, which are characterized by specific physical features. In other words, there is a direct correspondence between the $M$ coefficients and specific observable quantities. Hence, in order to reconstruct the dynamics of the perturbations we can chose to generate the evolution of the $M$ coefficients in the effective action or, equivalently, to work directly with the associated observable quantities. Let us therefore proceed in full generality: for any quantity described by a generic scalar function $Q$, we can always define a slow-roll parameter $s\q$ as follows:

\begin{equation}
s\q = - \dfrac{\dot{Q}}{H Q} = \dfrac{1}{\mpls} \dfrac{c(\phi)}{H(\phi) H'(\phi)} \dfrac{Q'(\phi)}{Q(\phi)}.
\end{equation}
With the same procedure as before, we can derive the hierarchy of $Q$-parameters: 

\begin{equation} \label{eq:Q_params_eft}
\begin{aligned} 
s\q(\phi) & = \dfrac{1}{\mpls} \dfrac{c(\phi)}{H(\phi) H'(\phi)} \dfrac{Q'(\phi)}{Q(\phi)}  \\[5pt]
\rho\q(\phi) & =  \dfrac{1}{\mpls} \, \dfrac{c(\phi)}{H'^{\,2}(\phi)} \dfrac{Q''(\phi)}{Q(\phi)}  \\
&\>\>\vdots \\
^{\,l}\chi\q(\phi) &= \biggl(\dfrac{c(\phi)}{\mpls}\biggl)^{l} \, \biggl( \dfrac{1}{H(\phi)} \biggl)^{l-1} \, 
\biggl( \dfrac{1}{H'(\phi)} \biggl)^{l+1} \, \dfrac{1}{Q(\phi)} \, \dfrac{d^{l+1}Q(\phi)}{d\phi^{l+1}} ,
\end{aligned}
\end{equation}
obeying the following Flow Equations:

\begin{equation} \label{eq:EFT_flow_eqs_Q}
\begin{cases}
\dfrac{ds \q}{dN} & =\quad  s \q(\theta - \epsilon - \eta - s \q) + \epsilon \rho\q \\[10pt]
\dfrac{d\rho\q}{dN}  &=\quad  \rho\q(\theta - 2\eta -s\q) +\,^{2}\chi\q \\
&\, \vdots   \\
\dfrac{d\,^{l}\chi\q}{dN}  &= \quad \,^{l}\chi\q \bigl[l \theta -(l-1) \epsilon) - (l+1)\eta - s\q \bigl] + \,^{l+1}\chi\q
\end{cases}
\end{equation}
where $l\geq2$ and $\rho\q$ is  $\,^{1}\chi\q$ and $\epsilon$,  $\eta$, $\theta$ are the parameters related to $H(\phi)$ and $c(\phi)$.

\subsubsection*{Fixed-points}
As a final step, we consider the overall system of equations displayed in Eqs.\eqref{eq:EFT_flow_eqs_bkg} and \eqref{eq:EFT_flow_eqs_Q} and we compute the fixed-points, for which all the derivatives vanish. By explicit calculation, it is possible to identify five different classes of fixed-points (notice that $l \geq 2$ in the following equations):

\begin{equation} 
\begin{aligned} 
& & &  \textrm{CLASS 1}:  \\[5pt]
& \epsilon = 0 \qquad &  & \theta = 0  \qquad & & s_{\textrm{\tiny{Q}}}  = 0 \qquad \\
& \eta = \textrm{const}  \qquad & & \kappa = \textrm{const}   \qquad&  &\rho_{\textrm{\tiny{Q}}} = \textrm{const} \qquad  \\
 \,^{l}&\lambda =  l! \, \eta^{l+1}  \qquad &  \,^{l}&\xi = l! \, \kappa \eta^{l-1}  \qquad & 
  \,^{l}  &\chi_{\textrm{\tiny{Q}}} =  l! \, \rho_{\textrm{\tiny{Q}}}\eta^{l-1}  \qquad
\end{aligned}
\bigskip
\end{equation}

\begin{equation} 
\begin{aligned} 
& & &  \textrm{CLASS 2}:  \\[5pt]
& \epsilon = 0 \qquad &  & \theta = 0  \qquad & & s_{\textrm{\tiny{Q}}}  = - \eta \\
& \eta = \textrm{const}  \qquad & & \kappa = \textrm{const}   \qquad&  &\rho_{\textrm{\tiny{Q}}} = \textrm{const}  \\
 \,^{l}&\lambda =  l! \, \eta^{l+1}  \qquad &  \,^{l}&\xi = l! \, \kappa \eta^{l-1}  \qquad & 
  \,^{l}  &\chi_{\textrm{\tiny{Q}}} =  (l-1)! \, \rho_{\textrm{\tiny{Q}}}\eta^{l-1} 
\end{aligned}
\bigskip
\end{equation}

\begin{equation} 
\begin{aligned} 
& & &   \textrm{CLASS 3}:  \\[5pt]
& \epsilon = 0 \qquad &  & \theta = \textrm{const}  \qquad & &s_{\textrm{\tiny{Q}}}  = 0 \\
& \eta = 0  \qquad & & \kappa = \textrm{const}   \qquad&  &\rho_{\textrm{\tiny{Q}}} = \textrm{const}  \\
 \,^{l}&\lambda = (-1)^{l-1} (l-1)! \, \theta^{l-1}  \qquad &  \,^{l}&\xi = 0 \qquad & 
  \,^{l}  &\chi_{\textrm{\tiny{Q}}} = (-1)^{l-1} (l-1)! \, \rho_{\textrm{\tiny{Q}}}\theta^{l-1} 
\end{aligned}
\bigskip
\end{equation}

\begin{equation} 
\begin{aligned} 
& & &  \textrm{CLASS 4}:  \\[5pt]
& \epsilon = 0 \qquad &  & \theta = \textrm{const}  \qquad & & s_{\textrm{\tiny{Q}}}  = \theta \\
& \eta = 0  \qquad & & \kappa = \textrm{const}   \qquad&  &\rho_{\textrm{\tiny{Q}}} = \textrm{const} \\
 \,^{l}&\lambda = (-1)^{l-1} (l-1)! \, \theta^{l-1}  \qquad &  \,^{l}&\xi = 0 \qquad & 
  \,^{l}  &\chi_{\textrm{\tiny{Q}}} = 0 \qquad \qquad \qquad \qquad \quad
\end{aligned}
\bigskip
\end{equation}

\begin{equation} 
\begin{aligned}
& \hspace{10em} \textrm{CLASS 5}:  \\[5pt]
& \begin{aligned} 
     & \epsilon = \textrm{const} \hspace{5em} & &  \theta = 2\epsilon   \\
     & \eta = \textrm{const} \hspace{5em} & & \kappa = \textrm{const}   \\
      \,^{l} & \lambda = \eta \prod_{m = 2}^{l} m\eta - (m-1)\epsilon  \hspace{5em}&   \,^{l} & \xi = 2(\epsilon+\eta) \prod_{m = 2}^{l} m\eta - (m-2)\epsilon 
    \end{aligned} \\[10pt]
& \begin{aligned}
       \hspace{7em} &s_{\textrm{\tiny{Q}}} (\epsilon -\eta - s_{\textrm{\tiny{Q}}}) + \epsilon \rho_{\textrm{\tiny{Q}}} = 0 \\
       \hspace{7em} &  \rho_{\textrm{\tiny{Q}}} (2\epsilon -2\eta - s_{\textrm{\tiny{Q}}}) + \,^{2}\chi_{\textrm{\tiny{Q}}} = 0 \\
       \hspace{7em} \,^{l}&\chi_{\textrm{\tiny{Q}}} =   \,^{l-1}  \chi_{\textrm{\tiny{Q}}} [s_{\textrm{\tiny{Q}}} - l(\epsilon-\eta)]
   \end{aligned}   
\end{aligned}
\end{equation}
Since there are so many parameters in play, a detailed analysis of the stability of these fixed-points similar to the one performed by Kinney in \citep{Kinney:2002qn} would not be particularly enlightening. Nevertheless, it is worth pointing out that attractive fixed points are preferred end-points of the evolutionary paths of the slow-roll parameters, whereas repulsive fixed points are naturally avoided. As we will see in the following, the structure of the fixed-points is responsible of the particular distribution of the inflationary models in the slow-roll parameter space.


\section{Observable predictions for the gravitational-wave sector}
\label{sec:observable_predictions}
The cosmological observables related to inflation can be expressed in terms of the slow-roll parameters. Therefore the power of the Flow Equations method is that it allows to work out observable predictions for a large number of stochastically generated models without relying on the specific form of their underlying actions. 
In the present work, we focus on the observables related to the tensor modes, namely the tensor-to-scalar ratio $r$ and the tensor spectral index $n_{t}$. In this section, we will derive a general expression of the consistency relation in the EFT framework, which will be used in order to study how the generated models populate the $(r, n_{t})$ plane. We will then briefly describe the algorithm that we developed for the integration of the Flow Equations. Finally, we will present the results of the analysis performed on $5 \times 10^{4}$ generated models. 

\subsection{Generalized consistency relation}
\label{subsec:consistency_rel}
The power-spectrum of scalar perturbations has already been computed in the EFT framework by solving a very general equation of motion for scalar fluctuations which reduces to the many inflationary models in the appropriate limits \citep{Bartolo:2010im,Bartolo:2010bj,Bartolo:2010di} (see also \citep{Ashoorioon:2018uey, Ashoorioon:2018ocr}). In particular, 
the expression that has been obtained retaining only the terms up to second order in the perturbations from the effective action is
\begin{equation} \label{eq:eft_scalar_power_spectrum}
\triangle _{\zeta}^{2} = \dfrac{(\alpha_{0}+ \sqrt{\beta_{0}})^{-\frac{3}{2}}H^{4}}{32 \pi \; \mpls \epsilon H^{2} \; \dfrac{1}{c^{\prime 2}_{s}}  \; \biggl| \Gamma \biggl( \dfrac{5}{4} +\dfrac{\alpha_{0}}{4\alpha_{0}-4i\sqrt{\beta_{0}}} \biggl) \biggl|^{2}},
\end{equation} 
where $\Gamma$ is the Euler Gamma function and $\alpha_{0}$, $\beta_{0}$ and $c^{\prime}_{s}$ are defined as follows:~\footnote{For an extension of the expressions for $\alpha_0$ and $\beta_0$ including some subleading terms from the action~(\ref{eq:action_com_gauge}), see Refs.~\cite{Bartolo:2010im,Bartolo:2010bj}.}
\begin{eqnarray} \label{eq:csp_a_b}
\alpha_{0} &\equiv& \dfrac{\mpls \epsilon H^{2} - \bar{M}_{1}^{3}H/2}{\mpls \epsilon H^{2} + 2M_{2}^{4}-3 \bar{M}_1^3 H}
\qquad
\beta_{0} \equiv \dfrac{(\bar{M}_{2}^{2} + \bar{M}_{3}^{2}) H^{2}}{2(\mpls \epsilon H^{2}+2M_{2}^{4} -3 \bar{M}_{1}^{3}H)}
\nonumber \\
c^{\prime 2}_{s} &\equiv&  \dfrac{ \mpls \epsilon H^{2}}{ \mpls \epsilon H^{2}+2M_{2}^{4}-3 \bar{M}_{1}^{3}H}\, .
\end{eqnarray}
The various quantities introduced in the expression above have a clear 
physical interpretation (see also, e.g.~\citep{Cheung:2007st,Bartolo:2010im}). The parameters $\alpha_0$ and $\beta_0$ determine the dispersion relation at horizon-crossing, 
which goes as $\omega^2 \simeq \alpha_0 k^2+\beta_0 k^4$ (leading to the specific dependence as $(\alpha_{0}+ \sqrt{\beta_{0}})^{-\frac{3}{2}}$, that accounts for the proper rescaling of momenta in the power-spectrum). The dispersion relation also determines the (phase and group) sound speed of scalar fluctuations. It turns out that at horizon crossing the group velocity of scalar fluctuations is $v_{g*} \simeq v_{p*}+(2 \beta_0/v_{p*}^3)$, where 
$v_{p*}^2=(1/2) ( \alpha_0+\sqrt{\alpha_0^2+8 \beta_0})$ is the phase velocity at horizon-crossing (for some details see Appendix~\ref{AppendixA}).
Notice that $\alpha_0$ reduces to the familiar expression of the sound speed $\alpha_0=v_p^2=c_s^2=(1-2M_2^4/\mpls \dot{H})^{-1}$ in the limit where $\bar{M}_{1}=0=\bar{M}_{0}$ (here $\bar{M}^2_{0}=
\bar{M}^2_{1}+\bar{M}^2_{2}$ and hence with $\beta_0=0$), corresponding to models of inflation with a generic dependence of the 
inflaton Lagrangian $\mathcal{L}=\mathcal{L}(\phi,X)$ on the kinetic term $X= - \frac{1}{2} g^{\mu \nu} \partial_\mu \phi 
\partial_\nu \phi$. The quantity that here we indicate with $c^{\prime 2}_{s}$ is related to the coefficient of the kinetic term $\dot{\pi}^2$ arising from the action~(\ref{eq:action_com_gauge}) (for the explicit expression of the action in terms of the Goldstone boson see Eq.\eqref{actionpi}), namely 
\begin{equation} 
\label{ktcsprime}
\mpls \epsilon H^{2}+2M_{2}^{4}-3 \bar{M}_{1}^{3}H=\mpls \epsilon H^{2} \frac{1}{c^{\prime 2}_{s}}\, .
\end{equation} 
Therefore, in the limit $\bar{M}_{1}=0=\bar{M}_{0}$ the parameter $c^{\prime 2}_{s}$ reduces to $c^{\prime 2}_{s}=\alpha_0=c_s^2$, leading to an expression for the scalar power-spectrum characteristic of $\mathcal{L}=\mathcal{L}(\phi,X)$ inflation models, namely $\triangle _{\zeta}^{2}=H^2/(8 \pi^2 \mpls \epsilon c_s)$.
However in general, $c^{\prime 2}_{s}$ is not a sound speed, and Eq.~(\ref{ktcsprime}), together with the expression for $\alpha_0$ in~(\ref{eq:csp_a_b}), makes it evident that when $\bar{M}_1$ is switched on, the branch with $\epsilon <0$ (NEC-violating models) opens up, corresponding to negative values of $c^{\prime 2}_{s}$ (which is required to guarantee the positiveness of the time kinetic term in order to avoid instabilities, in accordance also with other stability requirements discussed in Appendix~\ref{AppendixA}).~\footnote{Therefore $c^{\prime 2}_{s}$ can be considered also as a parameter associated to a non vanishing $\bar{M}_{1}$, measuring the deviation of $\alpha_0$ from its "usual" expression, given that we can also write $c^{\prime 2}_{s}=\alpha_0 (1-\bar{M}_{1}/ \epsilon \mpls H^2)$.}  Notice that to be conservative and avoid even for a short phase of the inflationary dynamics a temporary instability of the system, we have also imposed the stability condition $\alpha_0>0$. Therefore the way we can achieve $\epsilon <0$ is different from the case of the ghost condensate discussed in \citep{Creminelli:2006xe}. However our EFT expressions for the various quantities do include also this case, and it would not be difficult to apply our Flow equation method also to this possibility.

A similar computation within the EFT approach has been performed also for tensor perturbations \citep{Noumi:2014zqa,Creminelli:2014wna}. Taking into account all the operators in the quadratic effective action that induce tensor perturbations, one can obtain the following power-spectrum (at leading order in the slow-roll parameters)

\begin{equation} \label{eq:eft_tensor_power_spectrum}
\triangle \tens^{2}  =  c\tens^{-1}\dfrac{2 H^{2}}{\pi ^{2} \mpls}, 
\end{equation}
where $c \tens$ is the speed of sound of tensor modes. From this, we can derive the spectral index 

\begin{equation} \label{eq:eft_tensor_tilt}
n_{t} \equiv \dfrac{d\triangle\tens^{2}}{d\log k} 
\simeq \dfrac{1}{H} \dfrac{1}{\triangle\tens^{2}} \dfrac{d \triangle\tens^{2}}{dt}
= 2\dfrac{\dot{H}}{H^{2}} - \dfrac{\dot{c}\tens}{H c\tens} = - 2\epsilon + s\tens,
\end{equation}
where we defined a slow-roll parameter associated to $c \tens$:

\begin{equation}
s\tens = - \dfrac{\dot{c}\tens}{H c\tens}.
\end{equation}
The value and the sign of the tensor spectral index are therefore affected by the behaviour of both the Hubble parameter and the propagation speed of tensor modes. In particular, it has already been shown that a non-trivial evolution of $c \tens$ during inflation could lead to a blue tensor tilt (see e.g. \citep{Cai:2015yza, Cai:2016ldn}).
We can finally compute the tensor-to-scalar perturbation ratio $r = \triangle\tens^{2} / \triangle_{\zeta}^{2}$ dividing the tensor power-spectrum \eqref{eq:eft_tensor_power_spectrum} by the scalar power-spectrum \eqref{eq:eft_scalar_power_spectrum}. After some manipulations, we obtain

\begin{equation} \label{eq:eft_r}
r = \dfrac{64}{\pi} \dfrac{\epsilon}{c\tens c_{s}^{'2}} (\alpha_{0}+\sqrt{\beta_{0}})^{\frac{3}{2}}
\biggl| \Gamma \biggl(\dfrac{5}{4} + \dfrac{\alpha_{0}}{4\alpha_{0}-4i\sqrt{\beta_{0}}} \biggl)\biggl|^{2},
\end{equation}
which can be easily written as a function of $n_{t}$ using Eq.\eqref{eq:eft_tensor_tilt}. The final result for the consistency relation is therefore

\begin{equation} \label{eq:modified_consistency_relation}
r = - \dfrac{32}{\pi} \dfrac{n_{t}}{c \tens c_{s}^{'2}} (\alpha_{0}+\sqrt{\beta_{0}})^{\frac{3}{2}} |\Gamma |^{2} 
+ \dfrac{32}{\pi} \dfrac{s \tens}{c \tens c_{s}^{'2}} (\alpha_{0}+\sqrt{\beta_{0}})^{\frac{3}{2}} |\Gamma |^{2} \;,
\end{equation}
where we omitted the argument of the Euler Gamma function for simplicity. This consistency relation is very general and includes also the so-called \textit{super inflation} models, in which the Hubble parameter increases with time, so that $\epsilon <0$. 
As a final remark, we point out that Eq.\eqref{eq:modified_consistency_relation} reduces to the well-known consistency relations in the appropriate limits. For example, if we assume all the $M$ and $\bar{M}$ coefficients to be zero, we recover the results of standard single-field slow-roll inflation. In this limit, in fact, we have $\alpha_{0} = c_{s}^{\prime 2} = c \tens^{2} = 1$ and $ \beta_{0} = 0$, so that $s\tens =  0$, $|\Gamma ^{2}| = \pi / 4$ and \eqref{eq:modified_consistency_relation} reduces to $r = -8n_{t}$. If we consider instead only the $M_{2}$ term in the effective action, assuming that all the other $M$ and $\bar{M}$ parameters are negligible,
we obtain $\alpha_{0} = c_{s}^{\prime 2} \equiv c_{s}^2$, $c\tens = 1$ and $\beta_{0} = 0 $, so that $s\tens=  0$, $|\Gamma ^{2}| = \pi / 4$, thereby recovering the typical consistency relation for models with arbitrary speed of sound $r= -8c_{s}n_{t}$ \cite{Garriga:1999vw, Chen:2006nt}.

\subsection{The integration algorithm}
\label{subsec:algorithm}
In this section we will give a brief description of the algorithm that we implemented in order to integrate the Flow Equations that we discussed in the previous sections. 
Our algorithm is very similar to the original one designed by Kinney \citep{Kinney:2002qn} for the Hubble Flow Equations. We are therefore going to highlight the key points of its functioning and to explain the modifications that we brought in order to adapt the code to our system. We refer to some of the previous works about the Flow Equations for both standard single-field slow-roll inflation \citep{Peiris:2003ff,Ramirez:2005cy,Coone:2015fha} and inflationary models with non-standard kinetic term \citep{Peiris:2007gz,Powell:2008bi,Bessada:2009pe,Agarwal:2008ah} for more details. 
The task of the algorithm is to integrate the Flow Equations for the background \eqref{eq:EFT_flow_eqs_bkg} and for the physical quantities we are interested in \eqref{eq:EFT_flow_eqs_Q}. Since our goal is to plot the models in the $(r, n_{t})$ plane, we need to solve the equations for the quantities that appear in the consistency relation \eqref{eq:modified_consistency_relation}. As we mentioned, the system of equations \eqref{eq:EFT_flow_eqs_Q} can be specialized either to the $M_{i},\bar{M}_i$ parameters of the effective action or to the corresponding physical quantities, i.e. $c'_{s}$, $c \tens$, $\alpha_{0}$ and $\beta_{0}$. We decided to deal directly with the physical quantities, because in this way it is straightforward to establish suitable intervals for the initial conditions, as we will see in the following. For each model, the algorithm integrates numerically the systems \eqref{eq:EFT_flow_eqs_bkg} and \eqref{eq:EFT_flow_eqs_Q} of coupled differential equations truncated at a certain order M for a proper set of initial conditions, according to the following steps:
\begin{itemize}
\item Select a point in the slow-roll parameter space: $\epsilon_{0}$, $\eta_{0}$, ... , $\,^{\m}\lambda_{0}$; $\theta_{0}$, $\kappa_{0}$, ... , $\,^{\m}\xi_{0}$; $c^{\prime}_{s\,0}$, $s_{s\,0}$,  $\rho_{s\,0}$, ... , $\,^{\m}\chi_{s\,0}$ and the same for the truncated slow-roll hierarchy of $c\tens$, $\alpha_{0}$ and $\beta_{0}$.

\item Integrate forward in time (i.e. towards smaller values of $N$, the number of e-folds from the end of inflation) until either
\begin{itemize}
\item[\textit{a.}] the end of inflation, which is either given by the condition $\epsilon = 1$ for the models with $\epsilon > 0$ or, for models with $\epsilon <0$, forced by hand at a reference value $H_{\textrm{end}} < H_{\textrm{max}} = 2.7 \times 10^{- 5} \mpls$ ($95 \%$ CL) (see  \citep{Akrami:2018odb}).
\item[\textit{b.}] the evolution reaches a late-time fixed-point (only for the case $\epsilon > 0$).
\end{itemize}
\item If the late-time attractor is reached, evaluate the slow-roll parameters up to the M$^{\textrm{th}}$ order at that point.
\item If inflation ends, integrate the Flow Equations backwards in time up to $N_{\star} \simeq 60$ e-folds before the end of inflation, when the observable cosmological perturbations were produced (see \citep{Akrami:2018odb,Ade:2015lrj,Planck:2013jfk,Liddle:2003as} for further details), and calculate the parameters at that point. 
\item Reject all the models that violate the requirements of the inflationary paradigms or even the laws of fundamental physics. For instance, all the models that show ghosts, gradient instabilities or vacuum instabilities are rejected (see Appendix~\ref{AppendixA} for more details). Besides, all the models that produce a group velocity for scalar perturbations larger than one are non-causal and hence are rejected, as well as models that have $c \tens > 1$. Specifically we impose that the group velocity 
$v_{g*} \simeq v_{p*}+(2 \beta_0/v_{p*}^3) <1$ (where $v_{p*}^2=(1/2) (\alpha_0+\sqrt{\alpha_0^2+8 \beta_0})$). Moreover we have also imposed the condition $v^2_{g*} \gtrsim 4 \times 10^{-4}$ to satisfy the lower bound on the scalar sound speed obtained from the latest {\it Planck} constraints on primordial non-Gaussianity (for more details, see Appendix~\ref{AppendixA}).\footnote{ Notice that this justify our choice for the range of initial conditions of the parameter $\beta_0  \ \in \ [0,1/32]$ (in the case where $\alpha_0=0$) and of $\alpha_0 \ \in \ [0,1]$ (when $\beta_0=0$), see Eq.~(\ref{initialc}).}
We also reject all the models whose evolution do not support at least $N_{\star}$ e-folds. 
\end{itemize} 
With the help of this algorithm, we would like to try answering the question: what are the generic predictions of inflationary models encompassed by the EFT framework? In order to do this, we adopt a Monte Carlo approach to generate predictions for a large number of models, where each model corresponds to a randomly selected set of initial values for the slow-roll parameters up to second order. The initial conditions are chosen in the following ranges, which are related to the typical values of the involved quantities:

\begin{equation}
\label{initialc}
\begin{aligned}
\epsilon_{i} & \ \in \ 
\begin{cases}
\;[0,0.8] \; \textrm{for the case } \epsilon > 0   \\
\;[-0.8,0]  \; \textrm{for the case } \epsilon < 0
\end{cases}
\\
\eta_{i} & \ \in \ [-0.1,0.1] \\
^{2}\lambda_{i} & \ \in \ [-0.05,0.05] \\
& \; \\
\theta_{i} & \ \in \  [-0.1,0.1]
\\
\kappa_{i} & \ \in \ [-0.1,0.1] \\
^{2}\xi_{i} & \ \in \ [-0.05,0.05] \\
& \; \\
c_{si}^{\prime} & \ \in \ 
\begin{cases}
\;[0,10] \; \textrm{for the case } \epsilon > 0   \\
\;[-10,0]  \; \textrm{for the case } \epsilon < 0
\end{cases}
\\
s_{s i} & \ \in \ [-0.1,0.1] \\
\rho_{s i}& \ \in \ [-0.01,0.01] \\
& \; \\
c_{\textrm{\tiny{T}}i}, \alpha_{0i}  & \ \in \ [0,1] \\
s_{\textrm{\tiny{T}}i}, s_{\alpha i} & \ \in \ [-0.1,0.1] \\
\rho_{\textrm{\tiny{T}}i}, \rho_{\alpha i} & \ \in \ [-0.01,0.01] \\
& \; \\
\beta_{0i}  & \ \in \ [0,1/32] \\
s_{\beta i} & \ \in \ [-0.1,0.1] \\
\rho_{\beta i} & \ \in \ [-0.01,0.01] \\
\end{aligned}
\end{equation}
The exact choice of ranges for the initial values of the parameters does not have a large influence on the results, as long as they are chosen such that the slow-roll hierarchy is convergent (i.e. as long as the flow equations 
mantain a smooth ``slow-roll'' dynamics, with the various higher-order flow quantities that become higher-order in a generalized ``slow-roll'' hierarchy).  
For each repetition of the algorithm, the integration domain is taken to be of 100 e-folds. Finally, since its exact value is actually model-dependent, the number of e-folds from the end of inflation at which we evaluate the slow-roll parameters is randomly selected in the range
$N_{\star}  \ \in \ [40,70]$ for the $\epsilon > 0$ models and for the $\epsilon < 0$ models we run our simulations for $N_{\star}=40,50,60$. In both cases, as we will see, the final results are not sensibly affected by the exact choice of $N_{\star}$.

\subsection{Results: models in the \texorpdfstring{$(n_{t},r)$}{(nt,r)} plane}
\label{subsec:results_scatterplots}
By means of the algorithm described in the previous section, we obtained the various generalized slow-roll parameters evaluated at $N_{\star}$ e-folds before the end of inflation for a large number of simulated models. With these values in hand, we are able to compute the tensor-to-scalar ratio $r$~(\ref{eq:eft_r}) and the tensor spectral index $n_{t}$~(\ref{eq:eft_tensor_tilt}) for each model. As already mentioned, we analysed both inflationary models that respect to the Null-Energy Condition ($\epsilon > 0$) and models that violate it ($\epsilon < 0$), leading therefore to \textit{super inflation} \citep{Lucchin:1985wy}. We are now going to discuss the two scenarios separately, because they are treated in a slightly different way and their results are complementary.

\subsubsection*{EFT Inflation with \texorpdfstring{$\epsilon >0 $}{e>0}: red-tilted tensor spectra}
\begin{figure}
\centering
\includegraphics[width=1\textwidth]{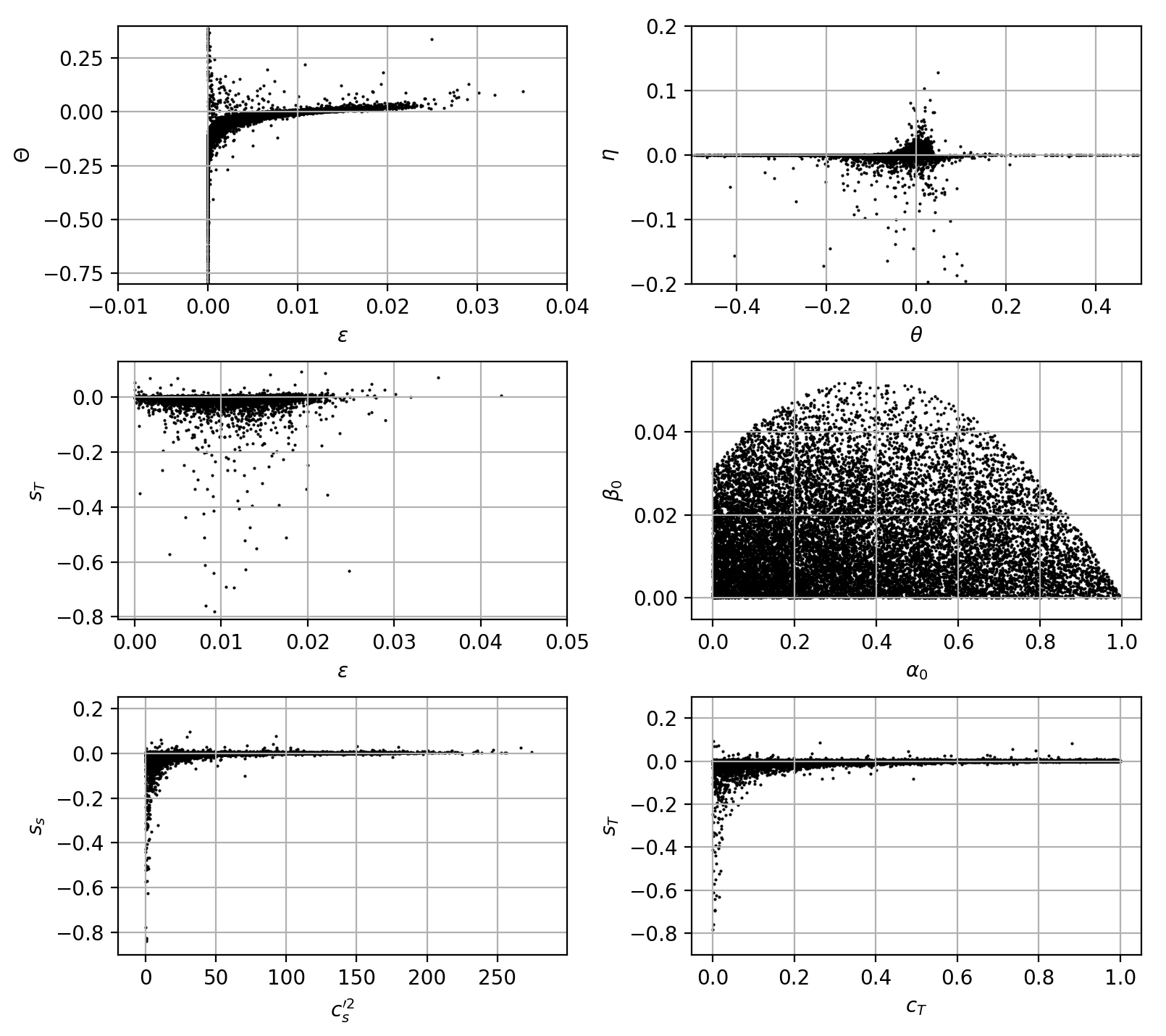}
\caption{Scatter-plots of $5 \times 10^{4}$ generated models in the $\epsilon > 0$ scenario (Null-Energy Condition respected during inflation). As an example, we have chosen to plot the models in the planes defined by some of the most relevant parameters. Due to the presence of fixed-points and to the causality controls that we impose in order to find the viable models, the points do not populate the plane uniformly but are rather clustered in particular regions.}
\label{fig:positive_eps_scatter_plots}
\end{figure}
We run our code for $5 \times 10^{4}$ inflationary models that respect the Null-Energy Condition. These models are characterized by $\dot{H} < 0$ and $\epsilon > 0$ and therefore the end of inflation is given by the condition $\epsilon = 1$. We rejected those models that did not respect the requirements of causality and of minimal duration of the inflationary epoch, which turned out to be around $70 \%$ of the total. The remaining models are considered viable and can be plotted in all the planes defined by the evolved parameters. As an example, in Figure \ref{fig:positive_eps_scatter_plots}, we show how the models populate the planes defined by some of the most relevant quantities. Remarkably, the points are not uniformly distributed but are rather clustered in particular regions of the planes. As originally pointed out in \citep{Kinney:2002qn}, this is partly due to the presence of many fixed-points in which the evolutionary paths of the slow-roll parameters are likely to be entrapped. The other factor that determines how the points are placed in the observable parameters space is given by the causality and minimum duration constraints that the models have to respect in order to be considered viable (see the description of the last step in the integration algorithm section~\ref{subsec:algorithm} above).
\begin{figure}
\centering
\includegraphics[width=1\textwidth]{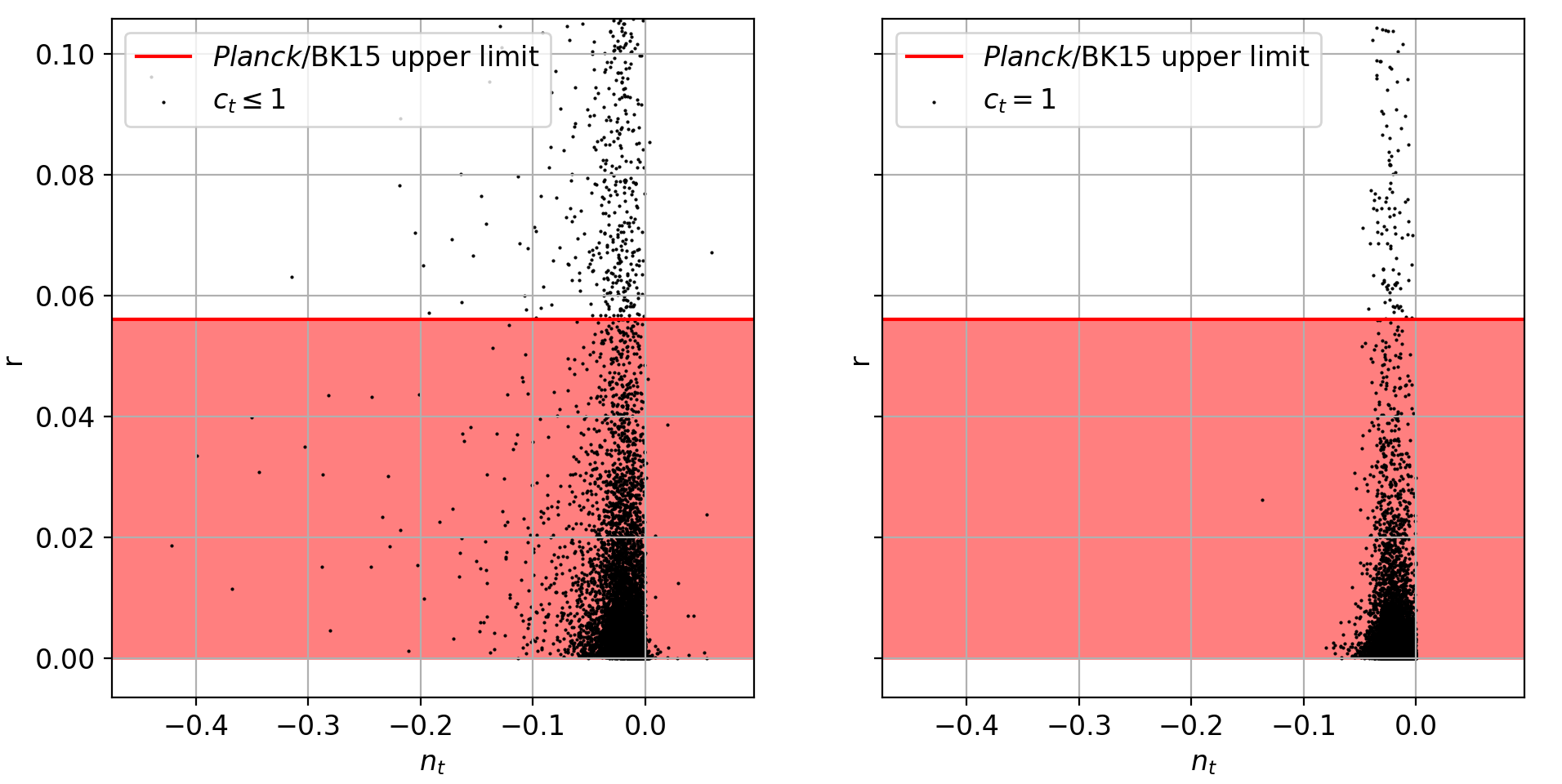}
\caption{Scatter-plot of $5 \times 10^{4}$ generated models in the plane $(n_{t},r)$ in the $\epsilon > 0$ scenario (Null-Energy Condition respected during inflation), for the general case  $c \tens \leq 1$ (left panel) and forcing $c \tens = 1$ (right panel). The values of $n_{t}$ and $r$ have been obtained substituting the output values of the algorithm inside-the general expressions \eqref{eq:eft_tensor_tilt} and \eqref{eq:modified_consistency_relation} respectively. The red shaded region is associated with the {\it Planck} upper bound for the tensor-to-scalar ratio: $r_{0.002} < 0.056$
($95 \%$ CL, Planck TT,TE,EE + lowE + lensing + BK15) \citep{Akrami:2018odb,Ade:2018gkx}. The figure shows only the portion of the plane which is not ruled out by observations: many of the original $5 \times 10^{4}$ points predicted too high values for $r$. Most of the models are characterized by a red-tilted tensor power-spectrum ($n_{t} < 0$) with a very mild slope ($n_{t} \gtrsim -0.06$).}
\label{fig:positive_eps_ntr}
\end{figure}
Interestingly enough, the models populate a large portion of the parameter space for $\alpha_0$ and $\beta_0$ that charaterize the disperion relation for the scalar fluctuations.
In order to study the general predictions of these models for what concerns the gravitational wave sector, we computed the tensor-to-scalar ratio and the tensor spectral index by means of Eqs.\eqref{eq:eft_tensor_tilt} and \eqref{eq:modified_consistency_relation}, respectively, and plotted the points in the $(r,n_{t})$ plane in Figure \ref{fig:positive_eps_ntr}, where we put in evidence all the points that lie in the portion of the plane that is not excluded by observations. The red shaded region is associated by the most recent upper bound on the value of the tensor-to-scalar ratio $r_{0.002} < 0.056$ ($95 \%$ CL, Planck TT,TE,EE + lowE + lensing + BK15), obtained by combining {\it Planck} and BICEP2/Keck Array BK15 data \citep{Akrami:2018odb,Ade:2018gkx}. A substantial portion of the viable models (around $90\%$) are below this upper limit and are therefore supported by observations.
Almost all of the points lie in the $n_{t} < 0$ half-plane, with only $\sim 1 \%$ of them having $n_{t} > 0$ and, in particular, most of those points show quite a flat tensor power-spectrum, with $n_{t} \gtrsim -0.06$. The reason why the vast majority of simulated models are characterized by a red-tilted tensor power-spectrum is probably a joint consequence of two facts. First, the impossibility of passing through $\epsilon = 0$ during the evolution, so that the first addend of $n_{t} = -2\epsilon - s \tens$ remains always negative. Second, the fact that the value of $s \tens$ seems to be always smaller that $\epsilon$. While the former fact has a clear physical meaning, the latter is probably due to the complexity of the Flow Equations system, characterized by many fixed-points. We recall that the expressions of $r$ and $n_{t}$ are such that $ n_{t} = -2\epsilon - s \tens$ and $r \propto \epsilon /c_{s}^{\prime 2}\, c \tens $. We therefore notice that the tensor-to-scalar ratio vanishes also in all fixed-points, except those of class 5, because $\epsilon = 0$. Moreover, we have that $n_{t}$ vanishes for the fixed-points of classes 1 and 3, which have also $s \tens = 0$. Finally, for the fixed-points of class 2 we have that $ n_{t} = -2\epsilon - s \tens \bigl|_{\textrm{\tiny{C2}}}  = \eta $, whereas for the fixed-points of class 4 we have $ n_{t} = -2\epsilon - s \tens \bigl|_{\textrm{\tiny{C4}}}  = -\theta$. We therefore deduce that for the fixed-points of classes 2 and 4 we have a blue tensor power-spectrum respectively for $\eta > 0 $ and $\theta < 0$. 
\newline
\indent
Finally, it is always possible to set to unity the speed of propagation of tensor modes during inflation by a suitable disformal transformation of the metric \citep{Creminelli:2014wna} (which would affect also the scalar perturbations).  For completeness, we run our code forcing $c \tens = 1$. The results are shown in the right panel of Figure \ref{fig:positive_eps_ntr}. We can notice that the models are slightly more clustered around the $r = n_{t} = 0$ scenario and that all the points distinctly lie in the $n_{t} < 0$ portion of the plane, without any outlier in the $n_{t} > 0$ region, as expected. Besides that, the results are substantially the same as the case for the general $c \tens \leq 1$.

In conclusion, even if chosen among the wide EFT zoology, inflationary models with $\epsilon > 0$ are characterized by red-tilted tensor power-spectra. This of course implies that there is less power on small scales, which makes such models very unlikely to be directly observable by gravitational-wave detectors in the near future.

\subsubsection*{EFT Inflation with \texorpdfstring{$\epsilon<0$}{e<0}: blue-tilted tensor spectra}
\begin{figure}
\centering
\includegraphics[width=1\textwidth]{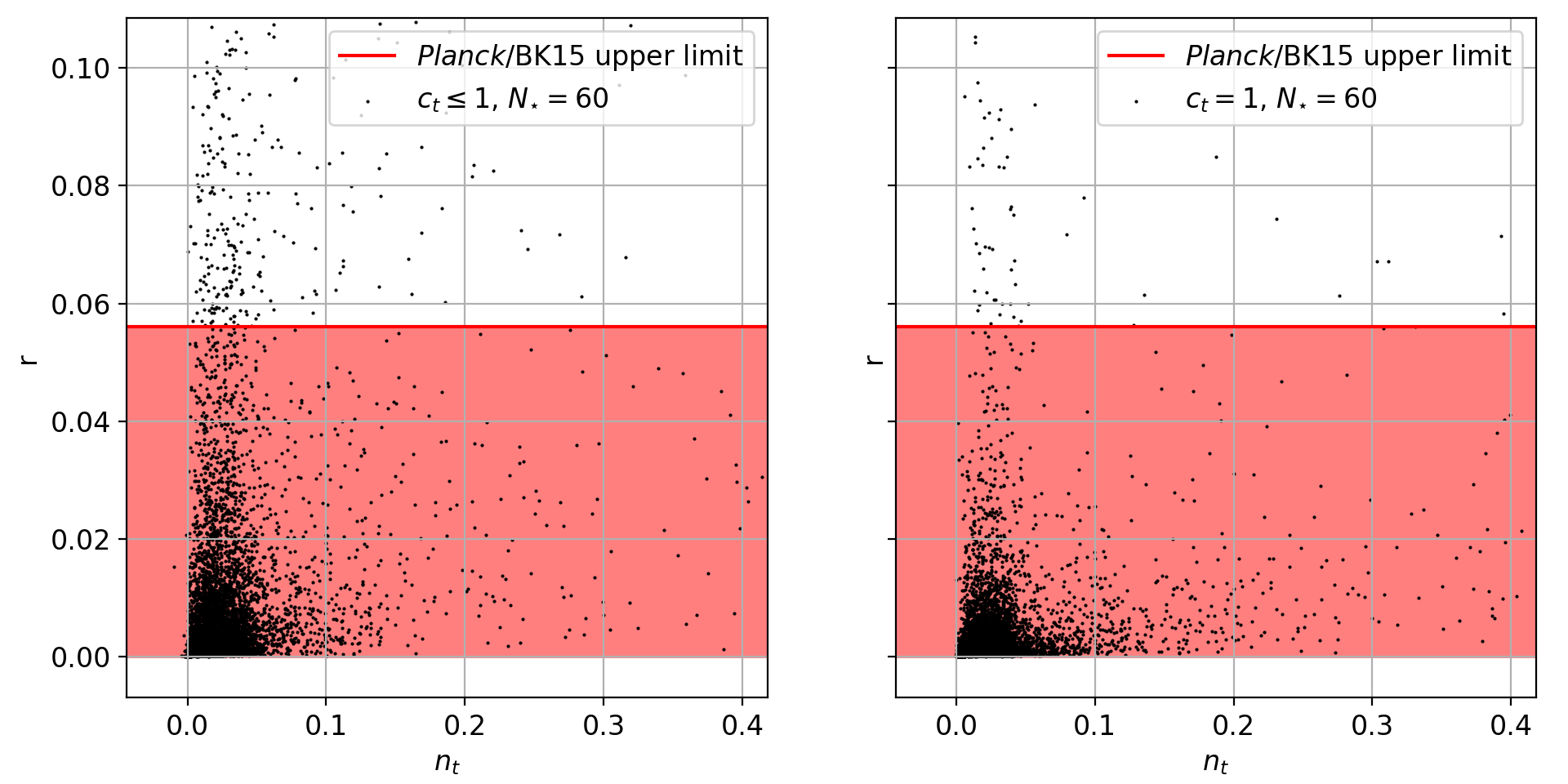}
\caption{Scatter-plot of $5 \times 10^{4}$ generated models in the plane $(n_{t},r)$ in the $\epsilon < 0$ scenario (super inflation models that violate the Null-Energy Condition), in the general case $c \tens \leq 1$ (left panel) and forcing $c \tens = 1$ (right panel). As for the $\epsilon > 0$ scenario, the value of $n_{t}$ has been obtained using the output values of the algorithm in the general expression \eqref{eq:eft_tensor_tilt}. The value of $r$, on the contrary, is obtained with the version of \eqref{eq:modified_consistency_relation} where $|\epsilon|$ is considered instead of $\epsilon$. The red shaded region is associated with the Planck upper bound for the tensor-to-scalar ratio: $r_{0.002} < 0.056$
($95 \%$ CL, Planck TT,TE,EE + lowE + lensing + BK15) \citep{Akrami:2018odb,Ade:2018gkx}. The figure shows only the portion of the plane which is not ruled out by observations. 
Interestingly, most of the models are characterized by a blue-tilted tensor power-spectrum ($n_{t} > 0$) with a mild slope ($n_{t} \lesssim 0.05$).}
\label{fig:negative_eps_ntr}
\end{figure}
\begin{figure}
\centering
\includegraphics[width=1\textwidth]{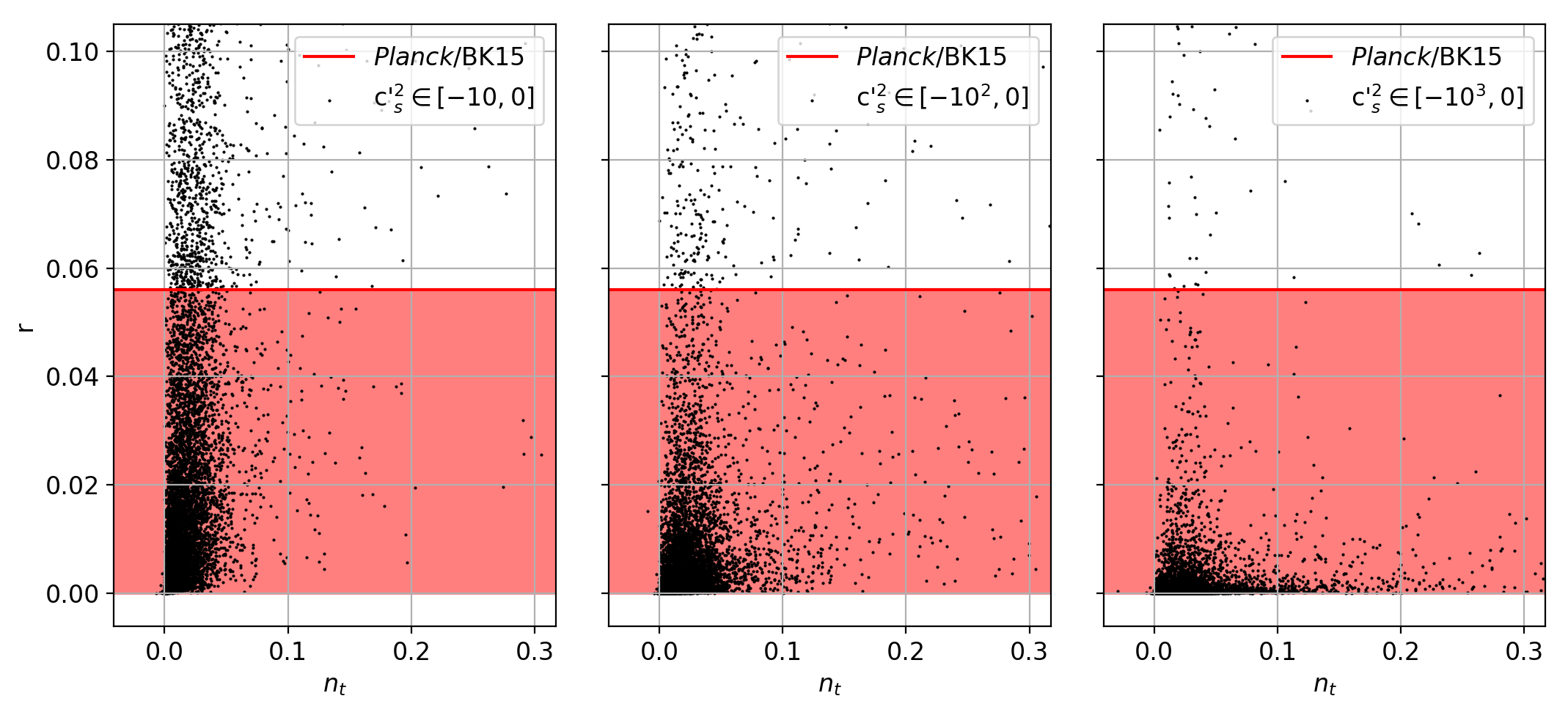}
\caption{Scatter-plot of $5\times10^{4}$ generated models in the plane $(n_{t},r)$ in the $\epsilon < 0$ scenario (super inflation models that violate the Null-Energy Condition) for three different choices for the initial range for $c^{\prime 2}_{s}$. As the range broadens the points shift towards smaller values of r and larger values of $n_{t}$. The former fact is trivially due to the dependence $r \propto 1/ c^{\prime 2}_{s}$ of the tensor-to-scalar ratio. The latter fact could be useful in order to characterize the models which are more likely to be detected with the Laser Interferometer Space Antenna (LISA) and the Einstein Telescope.}
\label{fig:negative_eps_diff_cs2p}
\end{figure}

We expect the situation to be quite different for super inflation models, which are known to be characterized by blue-tilted tensor power-spectra (see, e.g.,~\citep{Gasperini:1992pa,Brustein:1995ah,Baldi:2005gk,Creminelli:2006xe} and also~\citep{Ade:2015lrj} where {\it Planck} constraints have been obtained on Galileon inflation models). These models, as we already pointed out, are characterized by a negative $\epsilon$ and by a Hubble parameter which increases with time. Therefore, the end of inflation is no longer given by the condition 
$\epsilon = 1$, but instead is forced by external factors, typically by an additional field. Since the details of such a mechanism are not relevant for our purposes, we decided to set the end of inflation by hand at an arbitrary point of the evolution, taking into account the existing bounds on the end of inflation. We therefore included in our system of equations the one for the evolution of the Hubble parameter, 

\begin{equation}
\dfrac{\textrm{d}H}{\textrm{d}N} = H \epsilon \,,
\end{equation}
and we integrated it starting from a suitable initial condition $H_{0} \in [10^{-3}, 10^{14}]$ GeV. The range has been chosen so that the energy scale of inflation respects the upper bound on  $H_{\textrm{fin}} < 2.7 \times 10^{-5} \, \mpl$ \citep{Akrami:2018odb} and the lower limit around the MeV scale that guarantees hydrogen and helium production during Big Bang Nucleosynthesis \citep{Kawasaki:2000en,Giudice:2000ex,2004PhRvD..70d3506H,Katz:2016adq}. The Hubble parameter has been evolved forward in time for 100 e-folds, together with the other quantities and the slow-roll parameters. We designated the integration end-point as the end of inflation and we rejected all the realizations that did not respect the upper bound on the value of $H$, as well as the ones that did not support enough e-folds of inflation. We also rejected the models that are unstable or non-causal: see Appendix~\ref{AppendixA} for further details about the conditions we imposed in our code to identify and reject all the unhealthy models. Finally, we evaluated the observables quantities $N_{\star} = 60$ e-folds before the end of inflation for all the viable models, which turn out to be around $20 \%$ of the total.
The left panel of Figure \ref{fig:negative_eps_ntr} shows the results for $5 \times 10^{4}$ inflationary models in which the propagation speed of tensor modes is allowed to evolve. The vast majority of the generated models populate the region of the plane where $n_{t} > 0$, corresponding to blue-tilted tensor power-spectra. The points cluster around $n_{t} = 0$ and most of the models have $n_{t} < 0.06$, i.e. are highly scale-invariant. Nonetheless, the blue-tilted spectrum enhances the probability of a direct detection of the primordial gravitational-wave signal \citep{Gerbino:2014eqa,DiValentino:2011zz}. Indeed, even a small positive spectral index could produce an enhanced signal on small scales, given that the predicted power-spectra span $\sim 25$ orders of magnitude in the frequency domain \citep{Guzzetti:2016mkm,Ricciardone:2016ddg}. In recent works \citep{Bartolo:2016ami,Giare:2020vhn} it has been shown that in order to be detectable with the Laser Interferometer Space Antenna (LISA) and the Einstein Telescope the spectrum for inflationary gravitational waves must be considerably enhanced at scales smaller than the CMB ones. In particular, this requires a blue spectrum with $n_{t} \gtrsim 0.1-0.2$, which is almost one order of magnitude greater than the typical values that we are finding. In this regard, we noticed that changing the range for the initial conditions of the quantity $c^{\prime 2}_{s}$ defined in Eq.\eqref{eq:csp_a_b} we can obtain more models with a larger value of $n_{t}$. In Figure \ref{fig:negative_eps_diff_cs2p} we show the usual scatteplot in the $(n_{t},r)$ plane for three different choices of the initial range for $c^{\prime 2}_{s}$. We can notice that as the range broadens the points shift towards smaller values of r, which is expected since $r \propto 1/ c^{\prime 2}_{s}$, and larger values of $n_{t}$. We leave to future works a more detailed analysis of these issues, which could be useful in order to characterize the models which are more likely to be directly detected with next generation experiments.  
\newline
\indent
Finally, as we did for $\epsilon > 0$, we extended our analysis to the case where $c \tens=1$ during the entire evolution of the system, in order to test the stability of our results. Again, as we can see from the right panel of Figure \ref{fig:negative_eps_ntr}, we obtain that the plots are substantially unchanged, except from the fact that there are no more outliers in the $n_{t} < 0$ portion of the plane.
\begin{figure}
\centering
\includegraphics[width=1\textwidth]{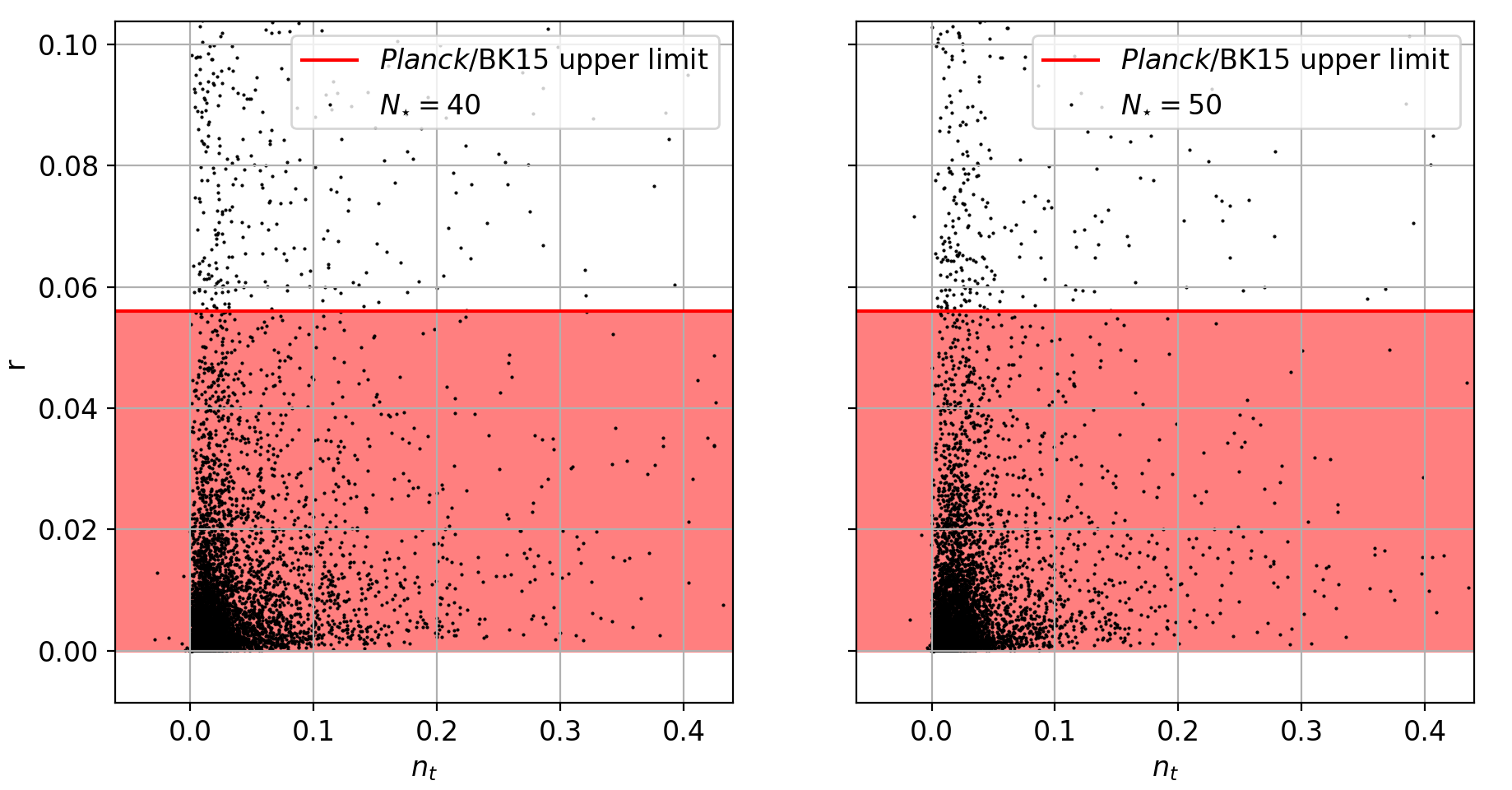}
\caption{Scatter-plot of $5\times10^{4}$ generated models in the plane $(n_{t},r)$ in the $\epsilon < 0$ scenario (super inflation models that violate the Null-Energy Condition) for two different choices of $N_{\star} = 40, 50$. The two plots are very similar to the one we showed in Figure \ref{fig:negative_eps_ntr}, which is obtained with $N_{\star} = 60$. This proves that our results do not depend strongly on the exact moment when the observable cosmological perturbations were produced. This is somehow reassuring, since the exact value of $N_{\star}$ is affected by the great uncertainties on the values of quantities such as the energy scale of inflation and the reheating temperature.}
\label{fig:negative_eps_diffN}
\end{figure}
As a further consistency check, we run the simulation for four different values of $N_{\star}$ in order to show that the final results do not depend strongly on the chosen value. In Figure \ref{fig:negative_eps_diffN} we show the scatter-plots in the $(n_{t}, r)$ plane of $5 \times 10^{4}$ points obtained with $N_{\star} = 40$ and $N_{\star} = 50$. These additional plots have to be compared with the results of Figure \ref{fig:negative_eps_ntr}, which are obtained with $N_{\star} = 60$ as we already mentioned. The three plots are quite similar and therefore we can conclude that our results are fairly independent of the choice of $N_{\star}$, whose exact value depends on still unknown quantities, such as the energy scale of inflation and the reheating temperature \citep{Akrami:2018odb}.


\section{Conclusions}
\label{sec:conclusion}
In this work we extended the method of Flow Equations to the EFT framework and we used it in order to generate a large number of beyond-standard inflationary models. We derived a general expression for the consistency relation valid for EFT models and we studied how the generated models populate the $(n_{t},r)$ plane. We found that a considerable fraction ($70 \%$) of the points are below the {\it Planck}/BK15 upper limit $r < 0.056$ \citep{Akrami:2018odb,Ade:2018gkx}. In the inflationary  scenario with $\epsilon > 0$ almost the entirety of points lie in the region $n_{t} < 0$, i.e. models with a red tensor power-spectrum are preferred. In the super inflation scenario ($\epsilon > 0$), on the other hand, the models are characterized by blue-tilted spectra ($n_{t} > 0$). We performed our analysis in the general case where the propagation speed of tensor modes  $c \tens$ can be different from one and allowed to vary during the evolution of the system and also in the case where it is forced to $c \tens = 1$: the main results are substantially the same in both cases.
The prediction of blue tensor spectra for NEC-violating inflationary scenarios is of great interest in view of a possible future detection of the stochastic background of primordial gravitational waves. Cosmological data are hopefully becoming precise enough to significantly constrain theoretical models of the early universe. For this reason, it will be worth making forecasts about the possibility of testing possible deviations from the standard consistency relation $r = -8 n_{t}$. This kind of analysis, together with the techniques described in this paper, could give important constraints on the physics of inflation and help to reduce the list of viable inflationary scenarios.


\appendix
\section{Appendix: EFT action and Stability conditions}
\label{AppendixA}
In this Appendix we provide some more details about the conditions that the EFT models should satisfy in order to avoid pathological instabilities of the system. These conditions have been accounted for and imposed to control 
the flow of the models and to eventually reject those outcomes that do not satisfy these conditions, as described in the integration algorithm section~\ref{subsec:algorithm}.  Some of these conditions have already been discussed~\cite{Cheung:2007st}, even though here we apply them to more general situations within the EFT models. 

It is well known that the leading-order terms of the (scalar part of the) quadratic EFT action in~(\ref{eq:action_com_gauge}) can be re-written as 
\begin{eqnarray}
\label{actionpi} 
S_\pi&=&\int  d^4x \sqrt{-g}\, \left[ \frac{1}{2} \mpls R  -\mpls \dot{H} \left( \dot{\pi}^2-\frac{(\partial_i \pi)^2}{a^2} \right) +2 M_2^4 \dot{\pi}^2-\bar{M}_1^3 H \left( 3\dot{\pi}^2-\frac{(\partial_i \pi)^2}{2a^2} \right) + \right. \nonumber \\ 
&-& \left. \frac{\bar{M}^2_2+\bar{M}_3^2}{2} \frac{(\partial^2_i \pi)^2}{a^4}  \right]\, ,
\end{eqnarray}
where one has re-introduced explicitly the scalar mode fluctuation $\pi$, related to the curvature perturbation by $\zeta=-H \pi$. 
As discussed in detail in Refs.~\cite{Bartolo:2010im,Bartolo:2010bj} (see also~\cite{Senatore:2004rj}) the equation of motion for the scalar degree of freedom $\pi$ is 
\begin{equation}
\label{eomu}
u''-\frac{2}{\tau^2} +\alpha_0 k^2 u+\beta_0 k^4 \tau^2 u=0\, , 
\end{equation}
where a prime denotes differentiation w.r.t. to conformal time $\tau$ and $u(k,\tau)= a(\tau) \pi(k,\tau)$. In Eq.~(\ref{eomu}) the coefficients $\alpha_0$ and $\beta_0$ are determined by the parameters $M_i$ and $\bar{M}_i$ of the action~(\ref{actionpi}), as given by the expressions~(\ref{eq:csp_a_b}) (for a detailed discussion about the quadratic action~(\ref{actionpi}), including sub-leading terms not shown in~(\ref{actionpi}) and their contribution to the expressions for $\alpha_0$ and $\beta_0$, see Refs.~\cite{Bartolo:2010im,Bartolo:2010bj}). 

The dispersion relation for the scalar fluctuations arising form the action~\ref{actionpi} is given by~\cite{Cheung:2007st,Bartolo:2010im}
\begin{equation}
\omega^2 = \alpha_0^2 k^2+\frac{\beta_0}{H^2} k^4 \, ,
\end{equation}
as can be seen also from the equation of motion~(\ref{eomu}). From here we read the phase and group velocity respectively
\begin{equation} 
\label{vpexplicit}
v_p=\frac{w}{k}=\sqrt{\alpha_0+\frac{\beta_0}{H^2}k^2}\, ,
\end{equation}
and 
\begin{equation}
v_g=\frac{dw}{dk}=\frac{v_p^2+(\beta_0/H^2)k^2}{v_p}\, .
\end{equation} 
We are mainly interested in evaluating these sound speeds at horizon crossing or, more precisely, when a given fluctuation mode gets frozen, namely when $\omega^2 \simeq 2 H^2$ (where the coefficient 2 is to adapt to the condition given by Eq.(2.6) of~\cite{Bartolo:2010im}), namely when 
\begin{equation}
\label{freezingMatte}
k^2 \simeq \frac{4 H^2}{(\alpha_0+\sqrt{\alpha^2_0+8 \beta_0})}\, ,
\end{equation}
where $k$ is a physical wavenumber. It follows that the group and phase velocity of scalar fluctuations at the freezing time $t_*$ are given by 
\begin{equation}
v_{g*}=v_{p*}+\frac{2\beta_0}{v^3_{p*}}\, , 
\end{equation} 
with 
\begin{equation} 
\label{vpf}
v^2_{p*}=\frac{1}{2} \left( \alpha_0+ \sqrt{\alpha^2_0+8 \beta_0}\right)\, .
\end{equation}
Notice that the result~(\ref{vpf}) is fully consistent with the expression that one can obtain for the phase velocity at freezing starting from $\omega^2 \simeq 2 H^2$ and then using $\omega^2 = v^2_{p} k^2$ and Eq.~(\ref{freezingMatte}), without using the explicit expression of $v_p$ in~(\ref{vpexplicit}).

To be conservative we have imposed the condition that the group velocity $v_{g*} <1$, in solving the flow equations for the EFT models, in order to avoid superluminal propagation. Again, it is not difficult to check that in terms of the "microscopic" parameters $M_i$ and $\bar{M}_i$  such a condition can be easily fulfilled. It turns out that, when $\beta_0$ is subdominant $v_g = v_p$ (as expected, since in this case the dispersion relation is $\omega \propto k$), while when $\beta_0$ dominates, then $v_g = 2 v_p$. This justifies our choice for the upper bounds of $\alpha_0$ and $\beta_0$ displayed in Eq.~(\ref{initialc}): 
$\alpha_0<1$ follows from the case when $\beta_0$ is subdominant, while the upper bound $\beta_0< (1/32)$ corresponds to $\beta_0$ dominating,  since in this case $v_{g*}= 2 v_{p*}=2 \left(\sqrt{8 \beta_0}\right)^{1/2}/ \sqrt{2}$. Finally, we have also imposed the condition $v^2_{g*} \gtrsim 4 \times 10^{-4}$ to satisfy the lower bound on the scalar sound speed derived from the latest {\it Planck} constraints on primordial non-Gaussianity. Even if such non-Gaussianity constraints actually refer to models where $\bar{M_i}=0$, we have nevertheless adopted them as the closest as possible reference constraint in this context. 

To avoid instabilities, one should require that the coefficient multiplying the kinetic term $\dot{\pi}^2$ in Eq.~(\ref{actionpi}) is positive
\begin{equation}
\label{c1}
\mpls \epsilon H^{2}+2M_{2}^{4}-3 \bar{M}_{1}^{3}H >0\, .
\end{equation}
To avoid gradient instabilities the coefficients of the spatial kinetic terms must be negative, therefore we need to require 
\begin{equation}
\label{c2}
\mpls \epsilon H^2- \bar{M}_{1}^{3}\frac{H}{2}>0\, , 
\end{equation}
and 
\begin{equation}
\label{c3}
\qquad
\bar{M}^2_0=(\bar{M}_{2}^{2} + \bar{M}_{3}^{2}) >0\, .
\end{equation}
It is not difficult to verify that the conditions~(\ref{c1}), (\ref{c2}) and~(\ref{c3}) can be easily satisfied in the parameter space of the "microscopic" parameters $M_i$ and $\bar{M}_i$ that switch on and off the various operators/models in the EFT action. In particular, we stress that all the previous inequalities can be satisfied also in the $\epsilon < 0$ case, since $\bar{M}_1$ can be negative.
Indeed, in our flow equation approach we have preferred to construct the hierarchy of the flow equations based on parameters, such as, e.g.,  
$\alpha_0$, $\beta_0$ and $c^{\prime 2}_{s}$ defined in Eq.~(\ref{eq:csp_a_b}), which have a more direct link with observational quantities. It is easy to check that the conditions~(\ref{c1}), (\ref{c2}) and~(\ref{c3}) are verified for $\alpha_0 > 0$ and $\beta_0 >0$, since the coefficient of the kinetic term is the denominator of $\alpha_0$ and $\beta_0$ and the coefficients of the spatial kinetic terms correspond to the numerators of $\alpha_0$ and $\beta_0$.  Consequently, we impose $\alpha_0 > 0$ and $\beta_0 >0$ in every moment during inflation as a condition for our models to be free of ghosts and gradient instabilities.
In the case where $M_i=0=\bar{M}_i$, the second condition~(\ref{c2}) necessarily implies that $\epsilon$ must be always positive. This is indeed the case of standard single-field models of inflation that is recovered in the limit 
$M_i=0=\bar{M}_i$. On the other hand, crucially the possibility to have $\epsilon <0$ (which can contribute to a blue spectral tilt of gravitational waves) is made possible thanks to a non vanishing 
$\bar{M}_1$, or equivalently, when the quantity $c^{\prime 2}_{s}<0$ (see Eq.~(\ref{ktcsprime})).  

As we have seen, an inflationary model that respects the conditions~(\ref{c1}), (\ref{c2}) and~(\ref{c3}) has no ghosts or gradient instabilities. However, it has been pointed out that for NEC violating models this may not be enough to ensure that the configuration is stable. In fact, the phase space of any NEC violating system (including those with no ghosts or gradient instabilities) has been shown to contain configurations with arbitrarily negative energy \citep{Sawicki:2012pz}. More precisely, the energy becomes negative only for observers boosted with a speed greater than a certain critical value (see Eq. (8) of \citep{Sawicki:2012pz}). To avoid such potentially unhealthy configurations, we impose that the group velocity of the scalar perturbations is smaller than the critical speed in any moment during inflation. As a final remark, we point out that G-inflation, which is free of all the instability issues discussed in \citep{Sawicki:2012pz}, is a subclass of the inflationary models that we study in our work. For a mapping of G-inflation in the EFT framework, see e.g. \citep{He:2016uiy, Burrage:2010cu}.

\acknowledgments

The authors acknowledge partial financial support by ASI Grant No. 2016-24- H.0. G.C. is supported by INFN initiative INDARK. We wish to thank Gianmassimo Tasinato and Matteo Fasiello for useful discussions.


\bibliographystyle{unsrt}  
\bibliography{flow_eft_revised}

\end{document}